\documentclass[3p,authoryear]{elsarticle}
\usepackage{tikz}
\usepackage[utf8]{inputenc}
\usetikzlibrary{decorations,arrows,shapes}
\usepackage{amsmath}
\usepackage[font=scriptsize,bf]{caption}
\newcommand{\final}{1}
\usepackage{fancyhdr}
\usepackage{verbatim}
\usepackage{pgfplots}
\usepackage{algorithm}
\usepackage{algpseudocode}
\usepackage{graphicx}
\PassOptionsToPackage{hyphens}{url}
\usepackage[colorlinks=true]{hyperref}

\makeatletter
\g@addto@macro{\UrlBreaks}{\UrlOrds}
\makeatother

\biboptions{square}

\algrenewcommand\algorithmicindent{1.0em}
\algtext*{EndWhile}	
\algtext*{EndIf}	
\algtext*{EndFor}	
\newcommand{\LineIf}[2]{\State\algorithmicif\ {#1}\ \algorithmicthen\ {#2}}
\newcommand{\D}[1]{}
\newcommand{\aand}{\textbf{and }}

\newcommand{\Continue}{\textbf{repeat loop}}

\fancypagestyle{plain}{
  \fancyhf{}
  \fancyhead[R]{\sc Priority-Flood}
  \cfoot{\thepage}
  
  }
\hypersetup{pdfauthor={Richard Barnes (ORCID: 0000-0002-0204-6040), Clarence Lehman, David Mulla},%
            pdftitle={Priority-Flood: An Optimal Depression-Filling and Watershed-Labeling Algorithm for Digital Elevation Models},%
            pdfproducer={LaTeX},%
            pdfcreator={pdfLaTeX}
}

\begin{document}
\thispagestyle{empty}

\noindent Cite as Barnes, Lehman, Mulla. ``Priority-Flood: An Optimal Depression-Filling and Watershed-Labeling Algorithm for Digital Elevation Models". Computers \& Geosciences. Vol 62, Jan 2014, pp 117--127. doi:  ``10.1016/j.cageo.2013.04.024".

\begin{frontmatter}
  \title{Priority-Flood: An Optimal Depression-Filling and Watershed-Labeling Algorithm for Digital Elevation Models}

	\author[rb]{Richard Barnes\corref{cor_rb}}
	\ead{rbarnes@umn.edu}
	\address[rb]{Ecology, Evolution, \& Behavior, University of Minnesota, USA}
	\cortext[cor_rb]{Corresponding author, 321-222-7637. ORCID: 0000-0002-0204-6040}

	\author[cl]{Clarence Lehman}
	\ead{lehman@umn.edu}
	\address[cl]{College of Biological Sciences, University of Minnesota, USA}

	\author[dm]{David Mulla}
	\ead{mulla003@umn.edu}
	\address[dm]{Soil, Water, and Climate, University of Minnesota, USA}

  \begin{abstract}
    \noindent Depressions (or pits) are low areas within a digital elevation model that are surrounded by higher terrain, with no outlet to lower areas. Filling them so they are level, as fluid would fill them if the terrain were impermeable, is often necessary in preprocessing DEMs. The depression-filling algorithm presented here---called Priority-Flood---unifies and improves on the work of a number of previous authors who have published similar algorithms. The algorithm operates by flooding DEMs inwards from their edges using a priority queue to determine the next cell to be flooded. The resultant DEM has no depressions or digital dams: every cell is guaranteed to drain. The algorithm is optimal for both integer and floating-point data, working in $O(n)$ and $O(n\log_2 n)$ time, respectively. It is shown that by using a plain queue to fill depressions once they have been found, an $O(m \log_2 m)$ time-complexity can be achieved, where $m$ does not exceed the number of cells $n$. This is the lowest time complexity of any known floating-point depression-filling algorithm. In testing, this improved variation of the algorithm performed up to 37\% faster than the original. Additionally, a parallel version of an older, but widely-used depression-filling algorithm required six parallel processors to achieve a run-time on par with what the newer algorithm's improved variation took on a single processor. The Priority-Flood Algorithm is simple to understand and implement: the included pseudocode is only 20 lines and the included \texttt{C++} reference implementation is under a hundred lines. The algorithm can work on irregular meshes as well as 4-, 6-, 8-, and $n$-connected grids. It can also be adapted to label watersheds and determine flow directions through either incremental elevation changes or depression carving. In the case of incremental elevation changes, the algorithm includes safety checks not present in prior works.
  \end{abstract}

\end{frontmatter}


\section{Background}
A digital elevation model (DEM) is a representation of terrain elevations above some common base level, usually stored as a rectangular array of floating-point or integer values. DEMs may be used to estimate a region's hydrologic and geomorphic properties, including soil moisture, terrain stability, erosive potential, rainfall retention, and stream power. Many algorithms for extracting these properties require (1)~that every cell within a DEM must have a defined flow direction and (2)~that by following flow directions from one cell to another, it is always possible to reach the edge of the DEM. These requirements are confounded by the presence of depressions and flats within the DEM.

Depressions (also known as pits) are inwardly-draining regions of the DEM which have no outlet. Sometimes representative of natural terrain, they may also result from technical issues in the DEM's collection and processing, such as from biased terrain reflectance or conversions from floating-point to integer precision.\citep{Nardi2008} A depression may be resolved either by breaching its wall (e.g. \citet{Garbrecht1998}), thus allowing it to drain to a nearby area of lower elevation, or by filling it.

DEMs have increased in resolution from thirty-plus meters in the recent past to the sub-meter resolutions becoming available today. Increasing resolution has led to increased data sizes: current data sets are on the order of gigabytes and increasing, with billions of data points. While computer processing and memory performance have increased appreciably during this time, legacy equipment and algorithms suited to manipulating smaller DEMs with coarser resolutions make processing these improved data sources costly, if not impossible. Therefore, improved algorithms are needed.

This paper presents an algorithm to resolve depressions by unifying and extending the work of several previous authors. Also presented are variants of this algorithm which can label watersheds and determine flow directions.

The general definition of the depression-filling problem was stated by \citet{Planchon2002}. Given a DEM $Z$, its depression-filled counterpart $W$ is defined by the following criteria:
\begin{enumerate}
	\item Each cell of $W$ is greater than or equal to its corresponding cell in $Z$.
	\item For each cell $c$ of $W$, there is a path that leads from $c$ to the boundary by moving downwards by an amount of at least $\epsilon$ between any two cells on the path, where $\epsilon$ may be zero. Such a path is referred to as an $\epsilon$-descending path.
	\item $W$ is the lowest surface allowed by properties (1) and (2).
\end{enumerate}
If $\epsilon=0$, then the third criterion is easy to achieve; however, if $\epsilon\ne0$, then special precautions must be taken, as described below.

The algorithm presented here is one of only two time-efficient algorithms for solving the depression-filling problem. Special cases of this algorithm have been described many times. These cases, their relations, and alternative algorithms are detailed below.

\section{Alternative Algorithms}
\citet{Arge2003} describes a specialized $O(n \log_2 n)$ procedure to perform watershed labeling, determine flow directions, and calculate flow accumulation on massive grids in situations where I/O must be minimized. Although parts of this procedure share sufficient similarities with Priority-Flood to be considered related, the combinations of algorithms used and the way in which they are specialized place it outside the scope of this paper. It is expected that the algorithm by Arge will run slower than that described here due to the greater overhead involved in explicit data management; however, this efficiency of memory may allow the algorithm by Arge to run better in situations where memory is limited.

There is also a widely-used algorithm for 8-connected grids by \citet{Planchon2002}. The algorithm works by flooding the entire DEM and then draining the edge cells. The entirety of the DEM is then repeatedly scanned to find the border of the drained and undrained regions; undrained cells adjacent to this border are then drained and increased in elevation by a small amount. The algorithm terminates when all cells have been drained.

\citet{Planchon2002} present two different implementations of their algorithm. The first is simple to implement, but inefficient, running in $O(n^{1.5})$ time. The second implementation runs in $O(n^{1.2})$ time during testing, but is much more complex, using 48 constants to define an iterative scan from multiple directions, a recursive upstream search with stack limiting to prevent overflows, and a quadruply-nested loop. The algorithm's design permits it to run in fixed memory. Unfortunately, the DEMs which require this are typically so large as to make running the Planchon--Darboux algorithm onerous.

Fortunately, a fast, simple, and versatile alternative algorithm is available. Here, it is referred to as the Priority-Flood Algorithm. Later, it will be shown that this alternative algorithm runs significantly faster than that by \citeauthor{Planchon2002}.

\begin{table*}
\centering
\scriptsize
\begin{tabular}{l l l l l l}
Year & Authors & Operation & Data Type & Connectedness & Time Complexity \\
\hline \\[-5pt] \vspace{5pt}
1989 & C.\ Ehlschlaeger  & \parbox{3cm}{\hbox{Flow Directions,} \hbox{Accumulation}} & Integer/Float* & 8, Any & $O(n^2)$* \\
1991 & Vincent \& Soille & Watershed Labels & Integer & Gridded, $n$-dim & $O(n)$ \\
1992 & S.\ Beucher \& F.\ Meyer & Watershed Labels & Integer & 4, 6, 8 & $O(n)$* \\
1994 & F.\ Meyer & Watershed Labels & Integer* & 8* & $O(n)$* \\
1994 & Soille \& Gratin & Filling & Integer & 4, 6, 8 & $O(n)$* \\
2006 & Wang \& H.\ Liu & Filling & Floating* & 8* & $O(n\log_2 n)$ \\
2009 & Y.H.\ Liu, Zhang, \& Xu & Filling & Floating* & 8 & $O(n\log_2 k)^\dagger$ \\
2010 & Metz, Mitasova, \& Harmon & Flow Directions & Floating* & 8* & $O(n\log_2 n)$* \\
2011 & Metz, Mitasova, \& Harmon & Flow Directions & Floating* & ``Gridded" & $O(n\log_2 n)$* \\
2011 & N.\ Beucher \& S.\ Beucher & Watershed Labels & Integer & 4, 6, 8 & $O(n)$* \vspace{5pt} \\
2012 & Magalh\~aes et.\ al & \parbox{3cm}{Filling, Flow Directions, Accumulation} & Integer & 8* & $O(n)$  \vspace{5pt} \\
2012 & Gomes et.\ al & \parbox{3cm}{Flow Directions, Accumulation} & Integer & 8* & $O(n)$
\end{tabular}
\caption{Summary of previous Priority-Flood variants. The table lists claims the authors have made. The (*) symbol indicates that the authors have not made a direct claim, so one has been inferred from their design choices. All the floating-point variants will also work on integer data, though the specified time complexities are then suboptimal. $^\dagger$\citet{Liu2009} claim a $O(8n\log_2 n)$ time complexity, but implement an $O(n\log_2 k)$ algorithm.}
\label{fig:previous_authors}
\end{table*}

\section{The Priority-Flood Algorithm}
\subsection{History}
In its most general form, the Priority-Flood Algorithm works by inserting the edge cells of a DEM into a priority-queue where they are ordered by increasing elevation. The cell with the lowest elevation is popped from the queue and manipulated. Following this, each neighbor which has not already been considered by the algorithm is manipulated and then added to the priority queue. The algorithm continues until the priority queue is empty.

The Priority-Flood Algorithm may be applied to either integer or floating-point DEMs and is optimal for both; the general algorithm is also indifferent as to the underlying connectedness of the DEM and works equally well on 4-, 6-, or 8-connected grids, as well as meshes. As detailed below, special cases of the Priority-Flood Algorithm have been independently described and improved by many authors. Table~\ref{fig:previous_authors} summarizes the work of these authors. The following is a historic overview of the Priority-Flood Algorithm followed by a description of the algorithm, an important improvement, and details of some of the algorithm's many variants.

\citet{Ehlschlaeger1989} was the first to suggest the Priority-Flood Algorithm, noting that
\begin{quote}
The most accurate method for determining watershed boundaries involves placing a person familiar with the nuances of contour maps at a drafting table to manually interpret drainage basins.
\end{quote}
He goes on to disparage the use of local 3\,x\,3 neighbourhoods in determining flow directions, pointing out that a manual interpreter would instead utilize a high-level view of the general flow of water and the location of drainage basins. His variation of the algorithm uses insertion sort and so has a sub-optimal average time complexity of $O(n^2)$ or more. Of all the authors mentioned here, \citeauthor{Ehlschlaeger1989} is the only one to mention modeling algorithms on human behavior.

\citet{Vincent1991} describe an $O(n)$ algorithm for labeling watersheds in integer digital images by sorting all the pixels into an ascending order. The image is then flooded upwards from its lowest pixels. A distance function is used to determine which cells form the boundaries between two watersheds.

\citet*{BeucherMeyer} present a similar $O(n)$ algorithm. Each local minima is assigned a unique label. The terrain is then flooded upwards from its lowest points using a hierarchical queue, with the result being a DEM wherein all cells which ultimately drain to a given minima bear that minima's label. \citeauthor{BeucherMeyer} provide a thorough discussion of hierarchical queues work and examples of their algorithm's application to image analysis.

\citet*{Meyer1994} gives an overview of many existing methods to derive watershed boundaries. He identifies plateaus and regions of equal elevation as being potentially problematic, but solves the problem by using first-in, first-out (FIFO) queues. This is analogous to the total order priority queues discussed later in this paper.

\citet*{Soille1994} were the first authors to apply Priority-Flood specifically to depression filling. Their $O(n)$ algorithm is described using the mathematics of morphological filters for use on 4-, 6-, or 8-connected grids with integer values. Their variation floods the DEM inwards from the edges and does not consider local minima.

\citet*{Wang2006} describe the first floating-point variant of the Priority-Flood Algorithm, identifying the time complexity in such a situation as being $O(n \log_2 n)$. They present a lucid comparison of their algorithm versus older, less-efficient depression-filling algorithms by \citet{Domingue1988}, \citet{Ocallaghan1984}, and \citet{Marks1984}.

\citet*{Liu2009} describe a variant of the algorithm by \citet{Wang2006}; whereas \citet{Wang2006} use a cell's presence in the priority queue as an indication of whether it should be processed, \citet{Liu2009} recognize that for an 8-connected grid a two-dimensional boolean array offers a clearer correlation between the pseudocode and an efficient implementation.

\citeauthor{Liu2009} claim a time complexity of $O(8n\log_2 n)$, but this is too high. Their implementation uses a sorted dictionary (i.e.\ a map) which bins all cells of equal elevation. As a result, their algorithm actually has a time of $O(n\log_2 k)$, where $k$ is the number of unique elevation levels, which may be less than $n$. As a result of this implementation detail and the small sizes of their test DEMs, \citeauthor{Liu2009} incorrectly conclude that Priority-Flood is slower than the Planchon--Darboux Algorithm.

\citet{Metz2010} and \citet{Metz2011} describe a variant of the algorithm using a totally-ordered priority queue (this will be explained below) to determine flow directions in watersheds. In the resulting DEM, every cell has a defined D8 flow direction which is guaranteed to drain to the edge of the DEM. \citeauthor{Metz2010} evaluate their method against a ``traditional sink filling method" and an impact reduction approach by \citet{Lindsay2005}; they find that their method yields flow accumulation streams which are closer to GPS field control points than the traditional method for all tested DEMs at all tested resolutions. Similarly, their algorithm out-performed the impact reduction approach in most cases. A detailed account of the method by \citeauthor{Metz2010} is given below.

\citet*{Beucher2011} describe an algorithm similar to that of \citet{BeucherMeyer}, but extend it to finding watershed boundaries and image reconstructions. The authors point out that for 32-bit integers a hierarchical queue may have to allocate impossibly large numbers of child queues. To counter this, they present space-efficient implementations of hierarchical queues and methods of overcoming this kind of over-allocation.

\citet{Magalhaes2012} describe a Priority-Flood variant using hierarchical queues to achieve $O(n)$ processing on an integer data set, and note that some floating-point data can be discretized. The algorithm they present is essentially the same as that described by \citet{Soille1994}, but flow directions are generated in the manner of \citet{Metz2010} and \citet{Metz2011}. Although using integer data limits the scope of their method, they claim it is a significant improvement over the floating-point algorithms presented by \citet{Wang2006} and \citet{Liu2009}.
\citet{Gomes2012} extends the work of \citet{Magalhaes2012} to increase the efficiency of the Priority-Flood Algorithm in situations where memory access must be limited.

\tikzset{%
xthin/.style={to path={%
\pgfextra{%
 \draw[line cap=round, line join=round, thin]
      (\tikztostart) -- (\tikztotarget);} (\tikztotarget) \tikztonodes}},
xthick/.style={to path={%
\pgfextra{%
 \draw[line cap=round, line join=round, thick]
      (\tikztostart) -- (\tikztotarget);} (\tikztotarget) \tikztonodes}},
xvthick/.style={to path={%
\pgfextra{%
 \draw[line cap=round, line join=round, ultra thick]
      (\tikztostart) -- (\tikztotarget);} (\tikztotarget) \tikztonodes}},
xultra thick/.style={to path={%
\pgfextra{%
 \draw[line cap=round, line join=round,ultra thick]
      (\tikztostart) -- (\tikztotarget);} (\tikztotarget) \tikztonodes}},
xthin/.style={to path={%
\pgfextra{%
 \draw[line cap=round, line join=round,thin]
       (\tikztostart) -- (\tikztotarget);} (\tikztotarget)  \tikztonodes}}}

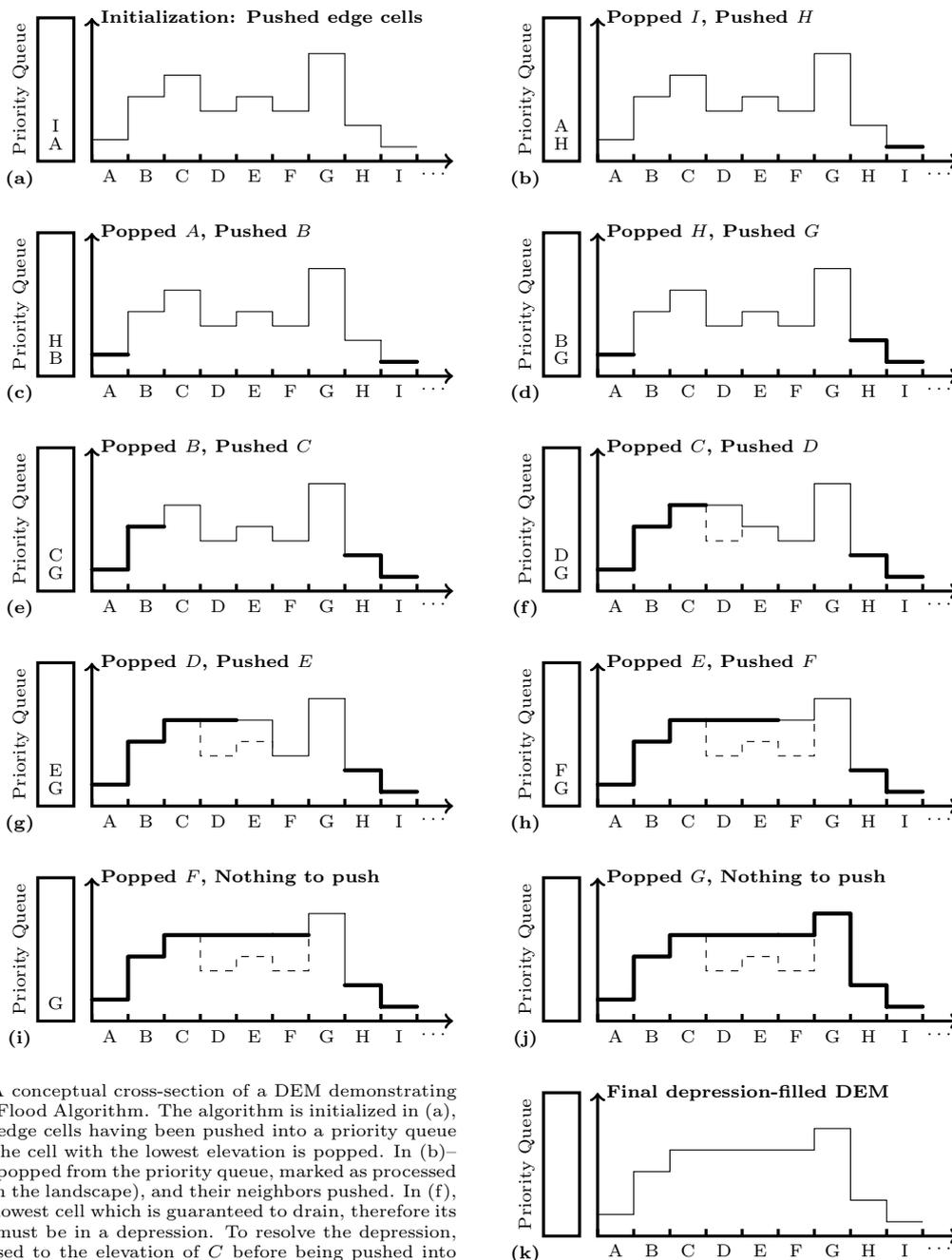
\begin{figure*}
\center
\begin{tikzpicture}
\tikzstyle{every node}=[font=\scriptsize]

\newcommand{\axes}[2]{
	\draw[<->,very thick] (0,2) -- (0,0)--(5,0);
	\foreach \x in {0,0.5,...,4.5}
	{
		\draw[very thick] (\x,0) -- (\x,0.1);
	}
	\draw[very thick] (-0.75,0) rectangle (-0.25,2);
	\node[rotate=90,above] at (-0.75,1) {Priority Queue};
	\node at (0.25,0)[below] {A};
	\node at (0.75,0)[below] {B};
	\node at (1.25,0)[below] {C};
	\node at (1.75,0)[below] {D};
	\node at (2.25,0)[below] {E};
	\node at (2.75,0)[below] {F};
	\node at (3.25,0)[below] {G};
	\node at (3.75,0)[below] {H};
	\node at (4.25,0)[below] {I};
	\node at (4.75,0)[below] {$\cdots$};
	\node at (-1,0)[below] {\textbf{(#1)}};
	\node at (0,2)[right] {\textbf{#2}};
}


\begin{scope}[shift={(0,0)}]
\axes{a}{Initialization: Pushed edge cells}
\draw (0,0.3) -- (0.5,0.3) -- (0.5,0.9) -- (1,0.9) -- (1,1.2) -- (1.5,1.2) -- (1.5,0.7) -- (2,0.7) -- (2,0.9) -- (2.5,0.9) -- (2.5,0.7) -- (3,0.7) -- (3,1.5) -- (3.5,1.5) -- (3.5,0.5) -- (4,0.5) -- (4,0.2) -- (4.5,0.2);
\node at (-0.5,0.50) {I};
\node at (-0.5,0.25) {A};
\end{scope}

\begin{scope}[shift={(7,0)}]
\axes{b}{Popped $I$, Pushed $H$}
\draw (0,0.3) -- (0.5,0.3) -- (0.5,0.9) -- (1,0.9) -- (1,1.2) -- (1.5,1.2) -- (1.5,0.7) -- (2,0.7) -- (2,0.9) -- (2.5,0.9) -- (2.5,0.7) -- (3,0.7) -- (3,1.5) -- (3.5,1.5) -- (3.5,0.5) -- (4,0.5) to[xthin] (4,0.2) to[xvthick] (4.5,0.2);
\node at (-0.5,0.50) {A};
\node at (-0.5,0.25) {H};
\end{scope}

\begin{scope}[shift={(0,-3)}]
\axes{c}{Popped $A$, Pushed $B$}
\draw (0,0.3) to[xvthick] (0.5,0.3) -- (0.5,0.9) -- (1,0.9) -- (1,1.2) -- (1.5,1.2) -- (1.5,0.7) -- (2,0.7) -- (2,0.9) -- (2.5,0.9) -- (2.5,0.7) -- (3,0.7) -- (3,1.5) -- (3.5,1.5) -- (3.5,0.5) -- (4,0.5) to[xthin] (4,0.2) to[xvthick] (4.5,0.2);
\node at (-0.5,0.50) {H};
\node at (-0.5,0.25) {B};
\end{scope}

\begin{scope}[shift={(7,-3)}]
\axes{d}{Popped $H$, Pushed $G$}
\draw (0,0.3) to[xvthick] (0.5,0.3) -- (0.5,0.9) -- (1,0.9) -- (1,1.2) -- (1.5,1.2) -- (1.5,0.7) -- (2,0.7) -- (2,0.9) -- (2.5,0.9) -- (2.5,0.7) -- (3,0.7) -- (3,1.5) -- (3.5,1.5) to[xthin] (3.5,0.5) -- (4,0.5) -- (4,0.2) to[xvthick] (4.5,0.2);
\node at (-0.5,0.50) {B};
\node at (-0.5,0.25) {G};
\end{scope}

\begin{scope}[shift={(0,-6)}]
\axes{e}{Popped $B$, Pushed $C$}
\draw (0,0.3) -- (0.5,0.3) -- (0.5,0.9) to[xvthick] (1,0.9) -- (1,1.2) -- (1.5,1.2) -- (1.5,0.7) -- (2,0.7) -- (2,0.9) -- (2.5,0.9) -- (2.5,0.7) -- (3,0.7) -- (3,1.5) -- (3.5,1.5) to[xthin] (3.5,0.5) -- (4,0.5) -- (4,0.2) to[xvthick] (4.5,0.2);
\node at (-0.5,0.50) {C};
\node at (-0.5,0.25) {G};
\end{scope}

\begin{scope}[shift={(7,-6)}]
\axes{f}{Popped $C$, Pushed $D$}
\draw (0,0.3) -- (0.5,0.3) -- (0.5,0.9) -- (1,0.9) -- (1,1.2) to[xvthick] (1.5,1.2) -- (2,1.2) -- (2,0.9) -- (2.5,0.9) -- (2.5,0.7) -- (3,0.7) -- (3,1.5) -- (3.5,1.5) to[xthin] (3.5,0.5) -- (4,0.5) -- (4,0.2) to[xvthick] (4.5,0.2);
\draw[dashed] (1.5,1.2) -- (1.5,0.7) -- (2,0.7) -- (2,0.9);
\node at (-0.5,0.50) {D};
\node at (-0.5,0.25) {G};
\end{scope}

\begin{scope}[shift={(0,-9)}]
\axes{g}{Popped $D$, Pushed $E$}
\draw (0,0.3) -- (0.5,0.3) -- (0.5,0.9) -- (1,0.9) -- (1,1.2) -- (1.5,1.2) to[xvthick] (2,1.2) -- (2.5,1.2) -- (2.5,0.7) -- (3,0.7) -- (3,1.5) -- (3.5,1.5) to[xthin] (3.5,0.5) -- (4,0.5) -- (4,0.2) to[xvthick] (4.5,0.2);
\draw[dashed] (1.5,1.2) -- (1.5,0.7) -- (2,0.7) -- (2,0.9) -- (2.5,0.9) -- (2.5,0.7);
\node at (-0.5,0.50) {E};
\node at (-0.5,0.25) {G};
\end{scope}

\begin{scope}[shift={(7,-9)}]
\axes{h}{Popped $E$, Pushed $F$}
\draw (0,0.3) -- (0.5,0.3) -- (0.5,0.9) -- (1,0.9) -- (1,1.2) -- (1.5,1.2) to[xvthick] (2.5,1.2) -- (3,1.2) -- (3,1.5) -- (3.5,1.5) to[xthin] (3.5,0.5) -- (4,0.5) -- (4,0.2) to[xvthick] (4.5,0.2);
\draw[dashed] (1.5,1.2) -- (1.5,0.7) -- (2,0.7) -- (2,0.9) -- (2.5,0.9) -- (2.5,0.7) -- (3,0.7) -- (3,1.2);
\node at (-0.5,0.50) {F};
\node at (-0.5,0.25) {G};
\end{scope}

\begin{scope}[shift={(0,-12)}]
\axes{i}{Popped $F$, Nothing to push}
\draw (0,0.3) -- (0.5,0.3) -- (0.5,0.9) -- (1,0.9) -- (1,1.2) -- (1.5,1.2) -- (2.5,1.2) to[xvthick] (3,1.2) -- (3,1.5) -- (3.5,1.5) to[xthin] (3.5,0.5) -- (4,0.5) -- (4,0.2) to[xvthick] (4.5,0.2);
\draw[dashed] (1.5,1.2) -- (1.5,0.7) -- (2,0.7) -- (2,0.9) -- (2.5,0.9) -- (2.5,0.7) -- (3,0.7) -- (3,1.2);
\node at (-0.5,0.25) {G};
\end{scope}

\begin{scope}[shift={(7,-12)}]
\axes{j}{Popped $G$, Nothing to push}
\draw (0,0.3) -- (0.5,0.3) -- (0.5,0.9) -- (1,0.9) -- (1,1.2) -- (1.5,1.2) -- (2.5,1.2) to[xvthick] (3,1.2) -- (3,1.5) -- (3.5,1.5) -- (3.5,0.5) -- (4,0.5) -- (4,0.2) to[xvthick] (4.5,0.2);
\draw[dashed] (1.5,1.2) -- (1.5,0.7) -- (2,0.7) -- (2,0.9) -- (2.5,0.9) -- (2.5,0.7) -- (3,0.7) -- (3,1.2);
\end{scope}

\begin{scope}[shift={(7,-15)}]
\axes{k}{Final depression-filled DEM}
\draw (0,0.3) -- (0.5,0.3) -- (0.5,0.9) -- (1,0.9) -- (1,1.2) -- (1.5,1.2) -- (3,1.2) -- (3,1.5) -- (3.5,1.5) -- (3.5,0.5) -- (4,0.5) -- (4,0.2) -- (4.5,0.2);
\end{scope}
\end{tikzpicture}

\vspace*{\dimexpr-\parskip-85pt\relax}
\caption{
\parshape 10
0pt .48\textwidth
0pt .48\textwidth
0pt .48\textwidth
0pt .48\textwidth
0pt .48\textwidth
0pt .48\textwidth
0pt .48\textwidth
0pt .48\textwidth
0pt .48\textwidth
0pt 1\textwidth
A conceptual cross-section of a DEM demonstrating the Priority-Flood Algorithm. The algorithm is initialized in (a), with all the edge cells having been pushed into a priority queue from which the cell with the lowest elevation is popped. In (b)--(e), cells are popped from the priority queue, marked as processed (dark lines on the landscape), and their neighbors pushed. In (f), cell $C$ is the lowest cell which is guaranteed to drain, therefore its neighbor $D$ must be in a depression. To resolve the depression, cell $D$ is raised to the elevation of $C$ before being pushed into the priority queue. $D$'s old elevation is no longer needed, but is shown with a dotted line for clarity. In (g), $E$ must also, by the same logic, be in a depression; therefore, it is raised to the new elevation of $D$. This continues through (i). In (j), the final cells are popped, though there is nothing to do. (k) shows the final elevations of the cells. Since the cells are raised as the algorithm progresses, no extra work is necessary between (j) and (k); the algorithm is complete when the priority queue is empty.}
\label{fig:swyfill}
\end{figure*}

\begin{algorithm}
\caption{{\sc Priority-Flood}: A generalization of the hierarchical-queue and priority-queue methods described by previous authors. The algorithm is described using pictures in Fig.~\ref{fig:swyfill}. \textbf{Upon entry,} \textbf{(1)}~\textit{DEM} contains the elevations of every cell or the value \textsc{NoData} for cells not part of the DEM. \textbf{(2)}~The value \textsc{NoData} is less than the elevation of any cell. \textbf{At exit,} \textbf{(1)}~\textit{DEM} contains the elevations of every cell or the value \textsc{NoData} for cells not part of the DEM. \textbf{(2)}~The elevations of \textit{DEM} are such that there are no digital dams and no undrainable depressions in the landscape, though there may be flats.}
\label{alg:swyfill}
\begin{algorithmic}[1]
	\Require \textit{DEM}
	\State Let \textit{Open} be a priority queue
	\State Let \textit{Closed} have the same dimensions as \textit{DEM}
	\State Let \textit{Closed} be initialized to {\sc false}
	\ForAll{$c$ on the edges of \textit{DEM}}
		\State Push $c$ onto \textit{Open} with priority \textit{DEM}($c$)
		\State $\textit{Closed}(c)\gets\textsc{true}$
	\EndFor
	\While{\textit{Open} is not empty}
		\State $c\gets\textsc{pop}(\textit{Open})$
		\ForAll{neighbors $n$ of $c$}
			\LineIf{$\textit{Closed}(n)$}{\Continue}
			\State $\textit{DEM}(n)\gets\textsc{max}(\textit{DEM}(n), \textit{DEM}(c))$
			\State $\textit{Closed}(n)\gets\textsc{true}$
			\State Push $n$ onto \textit{Open} with priority \textit{DEM}($n$)
		\EndFor
	\EndWhile
\end{algorithmic}
\end{algorithm}

\subsection{The Algorithm}
Generalizing the work of these authors yields the Priority-Flood Algorithm as described by Alg.~\ref{alg:swyfill} and Fig.~\ref{fig:swyfill}. To initialize the algorithm, all of the edge cells of the DEM (e.g.\ cells $A$ and $I$ in Fig.~\ref{fig:swyfill}a) are pushed (i.e.~added) onto a priority queue. The priority queue is ordered with cells of lower elevation having greater priority, so that the cell with the lowest elevation in the queue is always the cell popped (i.e.~removed from the queue for processing) first.

Furthermore, all of the edge cells are marked as resolved in a special boolean array. By definition, edge cells have an $\epsilon$-descending path to the DEM's edge and they are already at the correct elevation, so the depression-filling criteria are preserved.

DEMs---especially gridded DEMs---may be used to represent irregularly-shaped patches of landscape. These patches are akin to islands of data floating in a sea of cells which must be ignored. These ignorable cells are denoted with a special \textsc{NoData} value which is assumed to be some extremely negative number. Because this makes the \textsc{NoData} cells lower than any data cell, they have no impact on terrain flooding and can be treated as normal data cells.

Next, cells are popped off of the priority queue. Each cell $c$ which is popped is guaranteed to be the lowest cell with an established drainage path to the edge of the DEM; therefore, if $c$ is at the brink of a depression (such as cell $C$ is in Fig.~\ref{fig:swyfill}e), then there can be no lower cell which will drain the depression.

As each cell $c$ is popped from the priority queue, its neighbors are added to the priority queue, provided they have not already been resolved. If an unresolved neighbor $n$ is at a lower elevation than $c$ (such as cells $D,E,F$ in Fig.~\ref{fig:swyfill}), it is raised to the elevation of $c$ before it is placed on the queue. In this way, an $\epsilon$-descending path is constructed for each cell, leading to the fulfillment of the second depression-filling criterion. Because $n$ is always brought up only to the elevation of the lowest cell which still drains to the edge of the DEM, the third depression-filling criterion is fulfilled.

The algorithm terminates when the priority queue is empty (Fig.~\ref{fig:swyfill}k).

Since all cells are given elevations greater or equal to their original values in the DEM, the first property of the depression-filling criteria is fulfilled. Because all three criteria are fulfilled at each step, it follows that the result of the Priority-Flood Algorithm is a solution to the depression-filling problem.

Because the algorithm does not rely on the structure or connectedness of the underlying DEM, it can be applied to 4-, 6-, 8-, or, indeed, $n$-connected grids. The underlying grid need not even be rectangular---for instance, it may be a mesh based on a Delaunay triangulation.

\tikzset{%
xthin/.style={to path={%
\pgfextra{%
 \draw[line cap=round, line join=round, thin]
      (\tikztostart) -- (\tikztotarget);} (\tikztotarget) \tikztonodes}},
xthick/.style={to path={%
\pgfextra{%
 \draw[line cap=round, line join=round, thick]
      (\tikztostart) -- (\tikztotarget);} (\tikztotarget) \tikztonodes}},
xvthick/.style={to path={%
\pgfextra{%
 \draw[line cap=round, line join=round, ultra thick]
      (\tikztostart) -- (\tikztotarget);} (\tikztotarget) \tikztonodes}},
xultra thick/.style={to path={%
\pgfextra{%
 \draw[line cap=round, line join=round,ultra thick]
      (\tikztostart) -- (\tikztotarget);} (\tikztotarget) \tikztonodes}},
xthin/.style={to path={%
\pgfextra{%
 \draw[line cap=round, line join=round,thin]
       (\tikztostart) -- (\tikztotarget);} (\tikztotarget)  \tikztonodes}}}

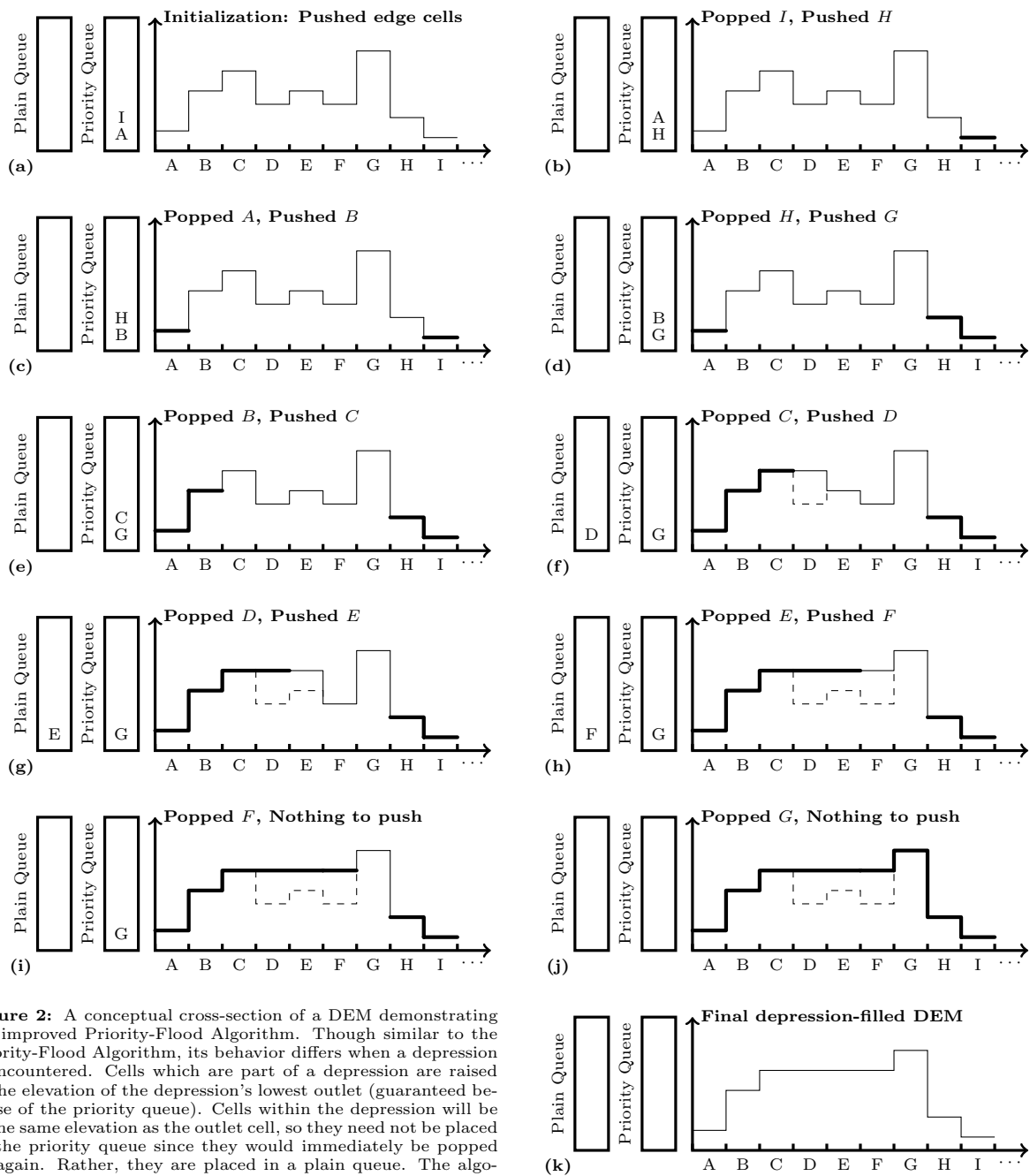
\begin{figure*}
\center
\begin{tikzpicture}
\tikzstyle{every node}=[font=\scriptsize]

\newcommand{\axes}[2]{
	\draw[<->,very thick] (0,2) -- (0,0)--(5,0);
	\foreach \x in {0,0.5,...,4.5}
	{
		\draw[very thick] (\x,0) -- (\x,0.1);
	}
	\draw[very thick] (-0.75,0) rectangle (-0.25,2);
	\node[rotate=90,above] at (-0.75,1) {Priority Queue};
	\draw[very thick] (-1.75,0) rectangle (-1.25,2);
	\node[rotate=90,above] at (-1.75,1) {Plain Queue};
	\node at (0.25,0)[below] {A};
	\node at (0.75,0)[below] {B};
	\node at (1.25,0)[below] {C};
	\node at (1.75,0)[below] {D};
	\node at (2.25,0)[below] {E};
	\node at (2.75,0)[below] {F};
	\node at (3.25,0)[below] {G};
	\node at (3.75,0)[below] {H};
	\node at (4.25,0)[below] {I};
	\node at (4.75,0)[below] {$\cdots$};
	\node at (-2,0)[below] {\textbf{(#1)}};
	\node at (0,2)[right] {\textbf{#2}};
}


\begin{scope}[shift={(0,0)}]
\axes{a}{Initialization: Pushed edge cells}
\draw (0,0.3) -- (0.5,0.3) -- (0.5,0.9) -- (1,0.9) -- (1,1.2) -- (1.5,1.2) -- (1.5,0.7) -- (2,0.7) -- (2,0.9) -- (2.5,0.9) -- (2.5,0.7) -- (3,0.7) -- (3,1.5) -- (3.5,1.5) -- (3.5,0.5) -- (4,0.5) -- (4,0.2) -- (4.5,0.2);
\node at (-0.5,0.50) {I};
\node at (-0.5,0.25) {A};
\end{scope}

\begin{scope}[shift={(8,0)}]
\axes{b}{Popped $I$, Pushed $H$}
\draw (0,0.3) -- (0.5,0.3) -- (0.5,0.9) -- (1,0.9) -- (1,1.2) -- (1.5,1.2) -- (1.5,0.7) -- (2,0.7) -- (2,0.9) -- (2.5,0.9) -- (2.5,0.7) -- (3,0.7) -- (3,1.5) -- (3.5,1.5) -- (3.5,0.5) -- (4,0.5) to[xthin] (4,0.2) to[xvthick] (4.5,0.2);
\node at (-0.5,0.50) {A};
\node at (-0.5,0.25) {H};
\end{scope}

\begin{scope}[shift={(0,-3)}]
\axes{c}{Popped $A$, Pushed $B$}
\draw (0,0.3) to[xvthick] (0.5,0.3) -- (0.5,0.9) -- (1,0.9) -- (1,1.2) -- (1.5,1.2) -- (1.5,0.7) -- (2,0.7) -- (2,0.9) -- (2.5,0.9) -- (2.5,0.7) -- (3,0.7) -- (3,1.5) -- (3.5,1.5) -- (3.5,0.5) -- (4,0.5) to[xthin] (4,0.2) to[xvthick] (4.5,0.2);
\node at (-0.5,0.50) {H};
\node at (-0.5,0.25) {B};
\end{scope}

\begin{scope}[shift={(8,-3)}]
\axes{d}{Popped $H$, Pushed $G$}
\draw (0,0.3) to[xvthick] (0.5,0.3) -- (0.5,0.9) -- (1,0.9) -- (1,1.2) -- (1.5,1.2) -- (1.5,0.7) -- (2,0.7) -- (2,0.9) -- (2.5,0.9) -- (2.5,0.7) -- (3,0.7) -- (3,1.5) -- (3.5,1.5) to[xthin] (3.5,0.5) -- (4,0.5) -- (4,0.2) to[xvthick] (4.5,0.2);
\node at (-0.5,0.50) {B};
\node at (-0.5,0.25) {G};
\end{scope}

\begin{scope}[shift={(0,-6)}]
\axes{e}{Popped $B$, Pushed $C$}
\draw (0,0.3) -- (0.5,0.3) -- (0.5,0.9) to[xvthick] (1,0.9) -- (1,1.2) -- (1.5,1.2) -- (1.5,0.7) -- (2,0.7) -- (2,0.9) -- (2.5,0.9) -- (2.5,0.7) -- (3,0.7) -- (3,1.5) -- (3.5,1.5) to[xthin] (3.5,0.5) -- (4,0.5) -- (4,0.2) to[xvthick] (4.5,0.2);
\node at (-0.5,0.50) {C};
\node at (-0.5,0.25) {G};
\end{scope}

\begin{scope}[shift={(8,-6)}]
\axes{f}{Popped $C$, Pushed $D$}
\draw (0,0.3) -- (0.5,0.3) -- (0.5,0.9) -- (1,0.9) -- (1,1.2) to[xvthick] (1.5,1.2) -- (2,1.2) -- (2,0.9) -- (2.5,0.9) -- (2.5,0.7) -- (3,0.7) -- (3,1.5) -- (3.5,1.5) to[xthin] (3.5,0.5) -- (4,0.5) -- (4,0.2) to[xvthick] (4.5,0.2);
\draw[dashed] (1.5,1.2) -- (1.5,0.7) -- (2,0.7) -- (2,0.9);
\node at (-1.5,0.25) {D};
\node at (-0.5,0.25) {G};
\end{scope}

\begin{scope}[shift={(0,-9)}]
\axes{g}{Popped $D$, Pushed $E$}
\draw (0,0.3) -- (0.5,0.3) -- (0.5,0.9) -- (1,0.9) -- (1,1.2) -- (1.5,1.2) to[xvthick] (2,1.2) -- (2.5,1.2) -- (2.5,0.7) -- (3,0.7) -- (3,1.5) -- (3.5,1.5) to[xthin] (3.5,0.5) -- (4,0.5) -- (4,0.2) to[xvthick] (4.5,0.2);
\draw[dashed] (1.5,1.2) -- (1.5,0.7) -- (2,0.7) -- (2,0.9) -- (2.5,0.9) -- (2.5,0.7);
\node at (-1.5,0.25) {E};
\node at (-0.5,0.25) {G};
\end{scope}

\begin{scope}[shift={(8,-9)}]
\axes{h}{Popped $E$, Pushed $F$}
\draw (0,0.3) -- (0.5,0.3) -- (0.5,0.9) -- (1,0.9) -- (1,1.2) -- (1.5,1.2) to[xvthick] (2.5,1.2) -- (3,1.2) -- (3,1.5) -- (3.5,1.5) to[xthin] (3.5,0.5) -- (4,0.5) -- (4,0.2) to[xvthick] (4.5,0.2);
\draw[dashed] (1.5,1.2) -- (1.5,0.7) -- (2,0.7) -- (2,0.9) -- (2.5,0.9) -- (2.5,0.7) -- (3,0.7) -- (3,1.2);
\node at (-1.5,0.25) {F};
\node at (-0.5,0.25) {G};
\end{scope}

\begin{scope}[shift={(0,-12)}]
\axes{i}{Popped $F$, Nothing to push}
\draw (0,0.3) -- (0.5,0.3) -- (0.5,0.9) -- (1,0.9) -- (1,1.2) -- (1.5,1.2) -- (2.5,1.2) to[xvthick] (3,1.2) -- (3,1.5) -- (3.5,1.5) to[xthin] (3.5,0.5) -- (4,0.5) -- (4,0.2) to[xvthick] (4.5,0.2);
\draw[dashed] (1.5,1.2) -- (1.5,0.7) -- (2,0.7) -- (2,0.9) -- (2.5,0.9) -- (2.5,0.7) -- (3,0.7) -- (3,1.2);
\node at (-0.5,0.25) {G};
\end{scope}

\begin{scope}[shift={(8,-12)}]
\axes{j}{Popped $G$, Nothing to push}
\draw (0,0.3) -- (0.5,0.3) -- (0.5,0.9) -- (1,0.9) -- (1,1.2) -- (1.5,1.2) -- (2.5,1.2) to[xvthick] (3,1.2) -- (3,1.5) -- (3.5,1.5) -- (3.5,0.5) -- (4,0.5) -- (4,0.2) to[xvthick] (4.5,0.2);
\draw[dashed] (1.5,1.2) -- (1.5,0.7) -- (2,0.7) -- (2,0.9) -- (2.5,0.9) -- (2.5,0.7) -- (3,0.7) -- (3,1.2);
\end{scope}

\begin{scope}[shift={(8,-15)}]
\axes{k}{Final depression-filled DEM}
\draw (0,0.3) -- (0.5,0.3) -- (0.5,0.9) -- (1,0.9) -- (1,1.2) -- (1.5,1.2) -- (3,1.2) -- (3,1.5) -- (3.5,1.5) -- (3.5,0.5) -- (4,0.5) -- (4,0.2) -- (4.5,0.2);
\end{scope}

\end{tikzpicture}
\vspace*{\dimexpr-\parskip-85pt\relax}
\caption{
\parshape 1
0pt .48\textwidth
A conceptual cross-section of a DEM demonstrating the improved Priority-Flood Algorithm. Though similar to the Priority-Flood Algorithm, its behavior differs when a depression is encountered. Cells which are part of a depression are raised to the elevation of the depression's lowest outlet (guaranteed because of the priority queue). Cells within the depression will be at the same elevation as the outlet cell, so they need not be placed on the priority queue since they would immediately be popped off again. Rather, they are placed in a plain queue. The algorithm is complete when both queues are empty. For integer data, this eliminates overhead; for floating-point data, this eliminates $O(\log_2 n)$ work per cell in the depression.}

\label{fig:barnesfill}
\end{figure*}

\subsection{An Important Improvement}
\label{sec:improve}
In the Priority-Flood Algorithm, cells are only raised when they are neighbors of the lowest cell which drains to the edge of the DEM. However, if a cell $c$ is raised, then it must have the same priority as the cell which raised it (this cell having had the highest priority in the priority queue); therefore, there is no need to push $c$ onto the priority queue and incur the associated cost. Rather, $c$ can be pushed onto a plain (first-in, first-out) queue at cost $O(1)$. 

This improvement is described by Alg.~\ref{alg:barnes1} and Fig.~\ref{fig:barnesfill}. As before, the algorithm is initialized by pushing all the edge cells of the DEM onto a priority queue (Fig.~\ref{fig:barnesfill}a). Cells are popped from this queue and their neighbors pushed onto it (Fig.~\ref{fig:barnesfill}b--e). However, if a neighbor is lower than the cell which is pushing it, the neighbor is raised to the elevation of that cell and added to a plain queue (Fig.~\ref{fig:barnesfill}f--i); if there are cells in the plain queue, they are always popped before cells in the priority queue. The algorithm terminates when there are no cells left in either queue (Fig.~\ref{fig:barnesfill}k).

\begin{algorithm}
\caption{{\sc Improved Priority-Flood}: This improvement to the {\sc Priority-Flood} uses a plain queue to speed-up the filling of depressions, once they are found. \textbf{Upon entry,} \textbf{(1)}~\textit{DEM} contains the elevations of every cell or the value \textsc{NoData} for cells not part of the DEM. \textbf{(2)}~The value \textsc{NoData} is less than the elevation of any cell. \textbf{At exit,} \textbf{(1)}~\textit{DEM} contains the elevations of every cell or the value \textsc{NoData} for cells not part of the DEM. \textbf{(2)}~The elevations of \textit{DEM} are such that there are no digital dams and no undrainable depressions in the landscape, though there may be flats.}
\label{alg:barnes1}
\begin{algorithmic}[1]
	\Require \textit{DEM}
	\State Let \textit{Open} be a priority queue
	\State Let \textit{Pit} be a plain queue
	\State Let \textit{Closed} have the same dimensions as \textit{DEM}
	\State Let \textit{Closed} be initialized to \textsc{false}
	\ForAll{$c$ on the edges of \textit{DEM}}
		\State Push $c$ onto \textit{Open} with priority \textit{DEM}($c$)
		\State $\textit{Closed}(c)\gets\textsc{true}$
	\EndFor
	\While{either \textit{Open} or \textit{Pit} is not empty}
		\If{\textit{Pit} is not empty}
			\State $c\gets\textsc{pop}(\textit{Pit})$
		\Else
			\State $c\gets\textsc{pop}(\textit{Open})$
		\EndIf
		\ForAll{neighbors $n$ of $c$}
			\LineIf{$\textit{Closed}(n)$}{\Continue}
			\State $\textit{Closed}(n)\gets\textsc{true}$
			\If{$\textit{DEM}(n)\le \textit{DEM}(c)$}
				\State $\textit{DEM}(n)\gets\textit{DEM}(c)$
				\State Push $n$ onto \textit{Pit}
			\Else
				\State Push $n$ onto \textit{Open} with priority \textit{DEM}($n$)
			\EndIf
		\EndFor
	\EndWhile
\end{algorithmic}
\end{algorithm}

\section{Ordering}
\label{sec:ordering}
A subtlety of priority queues concerns how elements of equal priority are treated. If elements of equal priority retain their order of insertion, then the priority queue may be said to be ``stable" or to have a ``total order". If elements of equal priority do not retain their order of insertion, then the priority queue has a ``strict weak ordering": elements of equal priority are incomparable and may therefore be arranged in any order.

Fortunately, for depression filling with $\epsilon=0$, it does not matter whether the priority queue has a total order or a strict weak order, as both will produce the same results. However, if $\epsilon\ne0$, or if Priority-Flood is used for watershed labeling or to determine flow directions directly, then the ordering is important and a total order is the best choice.

A total order produces a predictable, reproducible result, while the results of using an underlying algorithm with a strict weak order may only, in general, be reproduced by using that same algorithm. If $\epsilon\ne0$, then a total ordering guarantees the third depression-filling criterion; furthermore, a total order guarantees that each cell has a least-cost (i.e.\ shortest) path to its flooding source. This means that depressions are able to drain from multiple outlets, provided the outlets are of equal elevation. If the priority queue uses a strict weak ordering then it cannot make these guarantees.

In situations where a total order is necessary---e.g. $\epsilon\ne0$, watershed labeling, determining flow directions---it is possible to make an algorithm with strict weak ordering produce a total order. To do so, each cell is associated with two priorities (i.e.\ keys). The first is the cell's elevation, as before; the second is an integer which begins at zero and is incremented every time a cell is added to the priority queue. In the instance that two cells have equal elevation, the secondary priorities are used to determine the ordering. This assures that all cells are comparable and that cells of equal elevation emerge from the priority queue in the order they were inserted.

\section{Analysis}
\label{sec:analysis}
The essence of the Priority-Flood Algorithm is that a priority queue is initialized with some number of seed cells. The DEM is then progressively flooded by repeatedly pushing the neighbors of the highest priority cell into the priority queue. The improvemed Priority-Flood uses, in essence, a specialized priority queue.

As Table~\ref{tbl:pqueues} shows, there are many possible ways to manage a priority queue---and the list here is by no means exhaustive, \citet{Knuth1998} discusses many others in addition to presenting detailed analyses of some of the underlying priority queue algorithms. Yet, according to \citet{Edelkamp2011}, such management requires at most three functions. {\sc Insert} pushes an element onto a priority queue, {\sc DecreaseKey} deletes an arbitrary element and reinserts it into a more appropriate place, and {\sc DeleteMin} simultaneously accesses and pops the element with the greatest priority. The time complexity of operations on the priority queue is dependent on the time complexity of these operations.

\begin{table}
\centering
{\scriptsize
\begin{tabular}{l l l l}
\textbf{Data Structure}			& \textbf{Insert}			&	\textbf{DecreaseKey}					&	\textbf{DeleteMin}		\\
Hierarchical Queue$^\dagger$ & $O(1)$ & --- & $O(1)$ \\
Calendar Queue$^\ddagger$    & $O(1)$ & --- & $O(1)$ \\
1-Level Bucket			& $O(1)$			&	$O(1)$						&	$O(C)$					\\
2-Level Bucket			& $O(1)$			&	$O(1)$						&	$O(\sqrt{C})$			\\
Radix Heap				& $O(1)^*$			&	$O(1)^*$					&	$O(\lg C)^*$			\\
Emde Boas				& $O(\lg\lg N)$		&	$O(\lg \lg N)$				&	$O(\lg\lg N)$			\\
\\
Binary Search Tree		& $O(\lg n)$		&	$O(\lg n)$					&	$O(\lg n)$				\\
Binomial Queue			& $O(\lg n)$		&	$O(\lg n)$					&	$O(\lg n)$				\\
Heap					& $2\lg n$			&	$2 \lg n$					&	$2\lg n$				\\
Weak Heap				& $\lg n$			&	$\lg n$						&	$\lg n$					\\
Pairing Heap			& $O(1)$			&	$O(2^{\sqrt{\lg \lg n}})$	&	$O(\lg n)^*$			\\
Fibonacci Heap			& $O(1)$			&	$O(1)^*$					&	$O(\lg n)^*$			\\
Relaxed Weak Queue		& $O(1)$			&	$O(1)$						&	$O(\lg n)$				\\
\end{tabular}
}
\caption{Time complexities of priority queue operations for various underlying data structures---there are many other possibilities which are not listed. The top section of the table assumes that all priorities are integers. In the table, $C$ is the maximal edge weight, $N$ the maximum key, and $n$ the number of nodes stored. Values marked with ($^*$) denote amortized costs. The function `$\lg$' stands for `$\log_2$'. $^\dagger$As described by \citet{BeucherMeyer}. $^\ddagger$As described by \citet{Brown1988}, among other authors. All other entries are drawn from \citet{Edelkamp2011}.}
\label{tbl:pqueues}
\end{table}

For integer DEMs, it is evident that the priority queue should be based on one of the algorithms from the upper portion of Table~\ref{tbl:pqueues}. The algorithms from the lower portion will also work, albeit sub-optimally. In particular, hierarchical queues \citep{BeucherMeyer} are $O(1)$ with extremely low overhead for all operations; however, if there are many different elevation levels, the methods of \citet{Beucher2011} will need to be applied. Calendar queues \citep{Brown1988} have higher overhead, but this becomes negligible for large problems and they are well suited to the monotonically-increasing priorities exhibited by Priority-Flood. Calendar queues may be improved through delayed sorting \citep{Ronngren1991, Steinman1994, Steinman1996, Ronngren1997}, choosing or adaptively determining appropriately sized bins \citep{Oh1997, Oh1999, Tan2000, Hui2002, Siangsukone2003, Tang2005}, reducing resize operations \citep{Goh2003,Goh2004}, or leveraging statistical properties of the priority of newly-inserted elements \citep{Yan2006}. However, with the efficiency of the calendar queue comes subtleties in programming and determining algorithmic correctness. Bucket sorts or radix heaps may be simpler to implement while still retaining acceptable performance for most problems.

Regardless, since the priority queue operates in $O(1)$ time for integer data, Priority-Flood will operate in $O(n)$ time, which is optimal. In special cases, floating-point DEMs may be quantized and treated as integer DEMs. This may be the case when there are known resolution increments or if a floor function can be used to make suitably small buckets. Alternatively, an ``untidy priority queue" which rounds floating-point values may be used; \citet{Yatziv2006} describe such a queue and show that the error introduced by using it can be limited to the same order of magnitude as the errors introduced by spatial discretization, making them ``virtually insignificant".

In the general case of floating-point DEMs, an $O(\log_2 n)$ data structure must be used, such as one of those listed in the bottom portion of Table~\ref{tbl:pqueues}. The time complexity of Priority-Flood in such a scenario is $O(n \log_2 n)$. This may be reduced to $O(n \log_2 k)$ if the mapped queues method of \citet{Liu2009} is used.

However, time complexity alone does not tell the whole story: data structures have hidden overheads which are not expressible in terms of time complexity. \citet{Hendriks2010} performed extensive testing on implementations of eleven different priority queue algorithms and found sometimes dramatic differences in the time required to remove and then add an item to a filled queue. The algorithms also differed markedly for operations which filled an empty queue or emptied a filled queue.

Based on these tests, \citeauthor{Hendriks2010} recommended the \emph{implicit heap} as the best choice for a priority queue: it used the least memory of all the algorithms tested and operated the fastest, though it does use a strict weak ordering. Hierarchical heaps and ladder queues operated faster for very large data sets (greater than $10^6$ elements), but used significantly more memory.

For floating-point DEMs, the improved Priority-Flood above reduces both the time complexity---to $O(m \log_2 m)$, where $m\le n$---and run-time regardless of which underlying algorithm is used for the priority queue. It may also decrease run-times even for integer data by avoiding the overhead associated with maintaining a priority queue. If non-integer keys are used or guarantees regarding the internal structure of the priority queue are absent, prudent design suggests that the improved algorithm be used.

For all variants of the Priority-Flood Algorithm, the spatial distribution and connectedness of depressions in the DEM does not affect the algorithms' time complexities.

\section{Empirical Testing}
An implementation of the improved Priority-Flood was tested against an implementation of the unimproved Priority-Flood; both implementations are included in the Supplemental Materials. The \texttt{C++} STL priority queue was chosen to simplify programming and the tests were conducted on floating-point DEMs using a 64-bit Intel Xeon X5560 2.8GHz 8192kB cache processor and 24GB of RAM. The results are shown in Fig.~\ref{fig:wang_vs_me}. LIDAR-based 3\,m DEMs of 44~counties in Minnesota were tested, comprising an area of approximately 72,500\,km$^2$---one-third of the state---with approximately 111,111 cells per square kilometer for a total of approximately $8\cdot10^9$ cells; the average county size was 1,741\,km$^2$ ($2\cdot10^8$ cells). The improved Priority-Flood Algorithm out-performed the unimproved algorithm in every case. The average speed-up over all the counties tested was 16.8\%, while the median was 12.5\%. The maximal speed-up was 37.2\%.

\ifdefined\final
	\begin{figure}
	\begin{center}
	\begin{tikzpicture}[scale=0.75]
		\begin{axis}[ylabel=\% Speed-up of Improved vs. Unimproved, xlabel=\% of County Contained in Depressions, xmin=0, ymin=0]
		\addplot+[black, only marks] coordinates {
			(5.74053,2.75114) 
			(12.5404,3.51321) 
			(18.4719,5.80003) 
			(34.5639,4.26001) 
			(13.4422,4.95277) 
			(28.1587,5.83276) 
			(18.8461,6.70783) 
			(13.7239,6.85848) 
			(21.6795,7.78293) 
			(17.8287,8.66262) 
			(27.0075,8.7591) 
			(22.8042,10.0229) 
			(26.5512,10.0761) 
			(23.9933,10.4067) 
			(34.1044,10.5376) 
			(38.8416,11.1) 
			(27.5418,11.2269) 
			(26.502,11.3765) 
			(22.2307,11.8209) 
			(31.6474,12.2952) 
			(25.8206,12.3483) 
			(26.8477,12.3511) 
			(38.2473,12.6881) 
			(33.5834,12.954) 
			(46.8166,15.138) 
			(39.5356,15.4677) 
			(28.0958,17.5053) 
			(41.4193,18.6694) 
			(40.0249,20.2578) 
			(36.8583,24.468) 
			(45.0205,24.8552) 
			(39.4765,25.1687) 
			(51.7542,26.1133) 
			(49.7448,26.2809) 
			(57.6569,26.4934) 
			(44.1374,27.7961) 
			(48.7177,28.8888) 
			(48.6226,29.9859) 
			(58.1614,30.9802) 
			(57.6569,31.3222) 
			(53.836,31.5021) 
			(58.3566,31.8583) 
			(55.0003,34.1133) 
			(61.0962,37.2266) 
		};
		\end{axis}
	\end{tikzpicture}
	\end{center}
	\caption{Comparison of the improved Priority-Flood (Alg.~\ref{alg:barnes1}) and the unimproved Priority-Flood (Alg.~\ref{alg:swyfill}) algorithm. Tests were performed on 3\,m floating-point DEMs of 44 of Minnesota's counties, a 72,500$\,\textrm{km}^2$ area covering approximately 33\% of the state. It is clear that counties with more depressions are processed faster.}
	\label{fig:wang_vs_me}
	\end{figure}
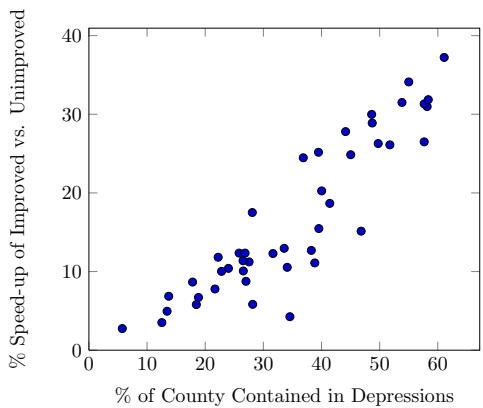
\fi



	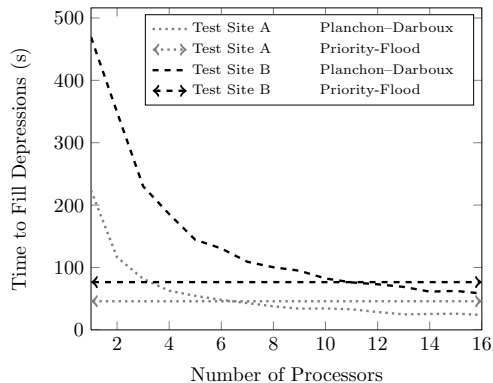
\begin{figure}
	\begin{center}
	\begin{tikzpicture}[scale=0.75]
		\begin{axis}[ylabel=Time to Fill Depressions (s), xlabel=Number of Processors, xmin=1, ymin=0, xmax=16]
		\addplot[gray,mark=none,dotted,very thick] coordinates{ 
			(1,223.701584)
			(2,116.401988)
			(3,81.910847)
			(4,62.811642)
			(5,54.583489)
			(6,48.136988)
			(7,43.003983)
			(8,37.827548)
			(9,33.961219)
			(10,34.426937)
			(11,33.036377)
			(12,28.636166)
			(13,24.925373)
			(14,25.526961)
			(15,26.031677)
			(16,23.972160)
		};
		\addplot[gray,<->,mark=none,domain=1:16,dotted, very thick] coordinates { (1,45.900834) (16,45.900834)}; 
		\addplot[black,mark=none,dashed,very thick] coordinates { 
			(1,469.205294)
			(2,348.759714)
			(3,230.186451)
			(4,185.590328)
			(5,144.471060)
			(6,130.571808)
			(7,109.207734)
			(8,100.409982)
			(9,94.856836)
			(10,82.439917)
			(11,76.170196)
			(12,73.141291)
			(13,68.902383)
			(14,61.344675)
			(15,62.496465)
			(16,58.332747)
		};
		\addplot[black,<->,mark=none,domain=1:16,dashed, very thick] coordinates { (1,76.586286) (16,76.586286)}; 
		\legend{\scriptsize\makebox[8em][l]{Test Site~A}\makebox[9em][l]{Planchon--Darboux},\scriptsize\makebox[8em][l]{Test Site~A}\makebox[9em][l]{Priority-Flood},\scriptsize\makebox[8em][l]{Test Site~B}\makebox[9em][l]{Planchon--Darboux},\scriptsize\makebox[8em][l]{Test Site~B}\makebox[9em][l]{Priority-Flood}};
		\end{axis}
	\end{tikzpicture}
	\end{center}
	\caption{Comparison of the TauDEM parallel implementation \citep{Wallis2009} of the Planchon--Darboux Algorithm versus the improved Priority-Flood algorithm. While the Planchon--Darboux Algorithm was tested with varying numbers of processors, the improved Priority-Flood was tested using only a single processor, resulting in the flat lines.}
	\label{fig:taudem_vs_me}
	\end{figure}

Earlier, the Planchon--Darboux Algorithm was discussed. This algorithm continues to see usage despite being slower than several Priority-Flood variants which have been developed both before and since. Notably, \citet{Wallis2009} produced an MPI-based parallel implementation of the Planchon--Darboux Algorithm for the TauDEM terrain analysis package and showed that it ran faster than a serial implementation for DEMs of moderate to large size. To demonstrate that the Priority-Flood Algorithm can run faster than the Planchon--Darboux Algorithm and to clarify a previous comparison made by \cite{Liu2009}, the two algorithms are compared here.

Both algorithms were compiled with full optimizations. The parallel implementation of Planchon--Darboux was run on varying numbers of processors while Priority-Flood was run on only a single processor. The machine used for the testing was the same as described above; it had eight processors per node and two dedicated nodes were used for parallel testing.

Due to long run-times, two test sites were chosen for comparing the speed of the improved Priority-Flood algorithm against the TauDEM 5.0.6 parallel implementation of the Planchon--Darboux algorithm. Test Site~A was Minnesota's Steele County and Test Site~B was Minnesota's Nicollet County. The two test sites represent counties with an average (35\%) and a large (61\%) number of depressions, respectively. Thus, the results of the test should not be biased by the choice of counties. The test sites were also representative in terms of size. Test Site A was a 3\,m DEM of 10891\,x\,13914 cells ($152\cdot10^6$ total) and Test Site B was a 3\,m DEM of 24140\,x\,13183 cells ($318\cdot10^6$ total).

As shown in Fig.~\ref{fig:taudem_vs_me}, the parallel implementation of the Planchon--Darboux Algorithm required six processors in order to process Test Site~A in the same time as the improved Priority-Flood algorithm did with one processor. The speed difference was even more striking in the case of Test Site~B, where 11 processors were required.

\begin{algorithm}
\caption{{\sc Priority-Flood+$\epsilon$}: This variation of the {\sc Improved Priority-Flood} ensures that all filled depressions have surfaces with a height difference of at least $\epsilon$ between any two consecutive cells as one moves away from or towards the depression's outlet. To minimize disruption to the DEM, Lines~\ref{alg:barnes1epseps1} and~\ref{alg:barnes1epseps2} use the \textsc{NextAfter} function, as described in the text on Page~\pageref{nextafter}. The algorithm raises a flag if the modified DEM has been altered in a way which cannot be minimized. \textbf{Upon entry,} \textbf{(1)}~\textit{DEM} contains the elevations of every cell or the value \textsc{NoData} for cells not part of the DEM. \textbf{At exit,} \textbf{(1)}~\textit{DEM} contains the elevations of every cell or the value \textsc{NoData} for cells not part of the DEM. \textbf{(2)}~The elevations of \textit{DEM} are such that there are no digital dams and every cell will drain to the edge of the DEM.}
\label{alg:barnes1eps}
\begin{algorithmic}[1]
	\Require \textit{DEM}
	\State Let \textit{Open} be a priority queue with total order\label{alg:barnes1epstotal}
	\State Let \textit{Pit} be a plain queue
	\State Let \textit{Closed} have the same dimensions as \textit{DEM}
	\State Let \textit{Closed} be initialized to \textsc{false}
	\ForAll{$c$ on the edges of \textit{DEM}}
		\State Push $c$ onto \textit{Open} with priority \textit{DEM}($c$)
		\State $\textit{Closed}(c)\gets\textsc{true}$
	\EndFor
	\While{either \textit{Open} or \textit{Pit} is not empty}
		\If{the top of \textit{Open} $=$ the top of \textit{Pit}}\label{alg:barnes1epstop1}
			\State $c\gets\textsc{pop}(\textit{Open})$\label{alg:barnes1epstop2}
      \State $\textit{PitTop}\gets\textsc{None}$\label{alg:pitreset1}
		\ElsIf{\textit{Pit} is not empty}
			\State $c\gets\textsc{pop}(\textit{Pit})$
      \If{$\textit{PitTop}=\textsc{None}$}
        \State $\textit{PitTop}\gets\textit{DEM}(c)$\label{alg:pitencounter}
      \EndIf
		\Else
			\State $c\gets\textsc{pop}(\textit{Open})$
      \State $\textit{PitTop}\gets\textsc{None}$\label{alg:pitreset2}
		\EndIf
		\ForAll{neighbors $n$ of $c$}
			\LineIf{$\textit{Closed}(n)$}{\Continue}
			\State $\textit{Closed}(n)\gets\textsc{true}$
			\If{$\textit{DEM}(n)=\textsc{NoData}$}
				\State Push $n$ onto \textit{Pit}
			\ElsIf{$\textit{DEM}(n)\le \textsc{NextAfter}(\textit{DEM}(c),\infty)$}\label{alg:barnes1epseps1}
        \If{$\textit{PitTop}<\textit{DEM}(n)$ \aand $\textsc{NextAfter}(\textit{DEM}(c),\infty)\ge\textit{DEM}(n)$}\label{alg:pitwarning}
          \State A significant alteration of the DEM has occurred
          \State The inside of the pit is now higher than the terrain surrounding it
        \EndIf
				\State $\textit{DEM}(n)\gets\textsc{NextAfter}(\textit{DEM}(c),\infty)$\label{alg:barnes1epseps2}
				\State Push $n$ onto \textit{Pit}
			\Else
				\State Push $n$ onto \textit{Open} with priority \textit{DEM}($n$)
			\EndIf
		\EndFor
	\EndWhile
\end{algorithmic}
\end{algorithm}

\section{Variants}
\subsection{Automatic Flat Resolution}
Priority-Flood, as it has been described above, will produce mathematically-flat surfaces as a by-product of filling depressions. Such surfaces confound attempts to determine flow directions and derivative hydrologic properties. A feature of the Planchon--Darboux Algorithm is that it does not suffer this problem because it produces surfaces for which each cell has a defined gradient from which flow directions can be determined.

As Alg.~\ref{alg:barnes1eps} demonstrates, Priority-Flood can be adapted to provide this functionality. To do so, depressions are filled such that adjacent cells along paths running to outlets are not set to the same elevation, but, rather, have some minimal elevation difference $\epsilon$ between themselves; that is, each cell is made to be a part of an $\epsilon$-descending path with $\epsilon>0$.

Although Alg.~\ref{alg:barnes1eps} is similar to Alg.~\ref{alg:barnes1}, there are important differences. Line~\ref{alg:barnes1epstotal} requires that the priority queue have a total order---as discussed in \textsection\ref{sec:ordering}. This allows depressions to drain from multiple outlets. Line~\ref{alg:barnes1epseps2} produces the $\epsilon$-descending path. The clause beginning on Line~\ref{alg:barnes1epstop1} maintains the total order by ensuring that all cells of equal elevation which may border a depression are treated equally.

Ideally, $\epsilon$ is large enough to direct flow, but sufficiently small as to have no other effects on the DEM's hydrologic properties. However, to work with such small values is to expose the intricacies of floating-point arithmetic. If too small an $\epsilon$ is used, then adding it to a cell may produce no change in its elevation; if too great an $\epsilon$ is used, then large depressions may be converted into mesas rising above the surrounding landscape. 

Choice of an appropriate value for $\epsilon$ is neither straight-forward nor universal. Therefore, it is best to use the \label{nextafter}\textsc{NextAfter} function defined by the \texttt{C99} and \texttt{POSIX} standards, or a similar function. This function increases or decreases a floating-point number by what is guaranteed to be the smallest possible increment in the direction of a second number. While this is better than using an arbitrarily-defined $\epsilon$, it is still not possible to guarantee that Alg.~\ref{alg:barnes1eps} will behave correctly in situations where a DEM's precision approaches that of its storage data type.

Therefore, the variable \textit{PitTop} is set (Line~\ref{alg:pitencounter}) whenever a new depression is encountered and reset (Lines~\ref{alg:pitreset1} and~\ref{alg:pitreset2}) after the depression has been resolved. If incrementing a cell causes it to rise above a cell which was previously higher than the level of the pit's outlet, as defined by \textit{PitTop}, then Line~\ref{alg:pitwarning} issues a warning. The Planchon--Darboux Algorithm does not provide such a safety check.

An alternative approach to the algorithm just described would be to use Priority-Flood to fill the DEM's depressions and a secondary algorithm to resolve the flat surfaces which result. \cite{BarnesFlats} provides an $O(n)$ algorithm for generating convergent flow patterns in such a situation. One could also impose flow directions on the DEM directly using a variant of the Priority-Flood Algorithm described below.

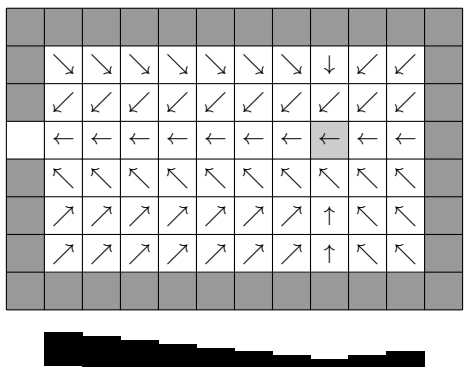
\begin{figure}
	\centering
	\begin{tikzpicture}
	\tikzstyle{every node}=[font=\footnotesize]
	\newcommand{\horizontallabels}[1]{
		\foreach \x in {1,...,12}{
			\node at (0.5*\x-0.25, #1) {\x};
		}
	}

	\newcommand{\verticallabels}[1]{
		\node at (#1,3.75) {A};
		\node at (#1,3.25) {B};
		\node at (#1,2.75) {C};
		\node at (#1,2.25) {D};
		\node at (#1,1.75) {E};
		\node at (#1,1.25) {F};
		\node at (#1,0.75) {G};
		\node at (#1,0.25) {H};
	}

	\newcommand{\gridcolor}{
		\fill [black!40] (0,0) rectangle (6,4);
		\fill [white] (0.5,0.5) rectangle (5.5,3.5);
		\fill [black!20] (4,2) rectangle (4.5,2.5);
		\fill [white] (0,2) rectangle (0.5,2.5);
		\draw[step=0.5cm,color=black] (0,0) grid (6,4);
	}

	\gridcolor{}

	\foreach \x in {0.5,1,...,4}
	{
		\fill [black] (\x,-0.75) rectangle (\x+0.51,-0.25-\x/10);
	}
		\fill [black] (4.5,-0.75) rectangle (5.01,-0.25-3.5/10);
		\fill [black] (5,-0.75) rectangle (5.51,-0.25-3/10);

	\node at (0.75,3.25) {$\searrow$};
	\node at (1.25,3.25) {$\searrow$};
	\node at (1.75,3.25) {$\searrow$};
	\node at (2.25,3.25) {$\searrow$};
	\node at (2.75,3.25) {$\searrow$};
	\node at (3.25,3.25) {$\searrow$};
	\node at (3.75,3.25) {$\searrow$};
	\node at (4.25,3.25) {$\downarrow$};
	\node at (4.75,3.25) {$\swarrow$};
	\node at (5.25,3.25) {$\swarrow$};
	\node at (0.75,2.75) {$\swarrow$};
	\node at (1.25,2.75) {$\swarrow$};
	\node at (1.75,2.75) {$\swarrow$};
	\node at (2.25,2.75) {$\swarrow$};
	\node at (2.75,2.75) {$\swarrow$};
	\node at (3.25,2.75) {$\swarrow$};
	\node at (3.75,2.75) {$\swarrow$};
	\node at (4.25,2.75) {$\swarrow$};
	\node at (4.75,2.75) {$\swarrow$};
	\node at (5.25,2.75) {$\swarrow$};
	\node at (0.75,2.25) {$\leftarrow$};
	\node at (1.25,2.25) {$\leftarrow$};
	\node at (1.75,2.25) {$\leftarrow$};
	\node at (2.25,2.25) {$\leftarrow$};
	\node at (2.75,2.25) {$\leftarrow$};
	\node at (3.25,2.25) {$\leftarrow$};
	\node at (3.75,2.25) {$\leftarrow$};
	\node at (4.25,2.25) {$\leftarrow$};
	\node at (4.75,2.25) {$\leftarrow$};
	\node at (5.25,2.25) {$\leftarrow$};
	\node at (0.75,1.75) {$\nwarrow$};
	\node at (1.25,1.75) {$\nwarrow$};
	\node at (1.75,1.75) {$\nwarrow$};
	\node at (2.25,1.75) {$\nwarrow$};
	\node at (2.75,1.75) {$\nwarrow$};
	\node at (3.25,1.75) {$\nwarrow$};
	\node at (3.75,1.75) {$\nwarrow$};
	\node at (4.25,1.75) {$\nwarrow$};
	\node at (4.75,1.75) {$\nwarrow$};
	\node at (5.25,1.75) {$\nwarrow$};
	\node at (0.75,1.25) {$\nearrow$};
	\node at (1.25,1.25) {$\nearrow$};
	\node at (1.75,1.25) {$\nearrow$};
	\node at (2.25,1.25) {$\nearrow$};
	\node at (2.75,1.25) {$\nearrow$};
	\node at (3.25,1.25) {$\nearrow$};
	\node at (3.75,1.25) {$\nearrow$};
	\node at (4.25,1.25) {$\uparrow$};
	\node at (4.75,1.25) {$\nwarrow$};
	\node at (5.25,1.25) {$\nwarrow$};
	\node at (0.75,0.75) {$\nearrow$};
	\node at (1.25,0.75) {$\nearrow$};
	\node at (1.75,0.75) {$\nearrow$};
	\node at (2.25,0.75) {$\nearrow$};
	\node at (2.75,0.75) {$\nearrow$};
	\node at (3.25,0.75) {$\nearrow$};
	\node at (3.75,0.75) {$\nearrow$};
	\node at (4.25,0.75) {$\uparrow$};
	\node at (4.75,0.75) {$\nwarrow$};
	\node at (5.25,0.75) {$\nwarrow$};

	\end{tikzpicture}
	\caption{Demonstration of Priority-Flood+FlowDirs (Alg.~\ref{alg:pf_flowdirs}). The dark shaded edge cells represent the border of a depression: their flow directions are omitted. The depression has an outlet along its left-hand side and a lightly shaded sink cell to the right of its center. The cells of each column---except for the sink, which is lower than any cell---have a relative elevation corresponding to the black elevation profile. }
	\label{fig:pf_flowdirs}
\end{figure}

\begin{algorithm}
\caption{{\sc Priority-Flood+FlowDirs}: This variation of {\sc Priority-Flood} is based on the work of \citet{Metz2011} and determines flow directions for all cells. This is, in effect, a depression-carving operation. On 8-connected grids, non-diagonal neighbors should be processed first on Line~\ref{alg:pf_flowdirs_ndn} as they have the greatest center-to-center slopes and therefore should take priority. A plain queue cannot be used to speed this variation up because all the elevation information inside the depressions is needed. \textbf{Upon entry,} \textbf{(1)}~\textit{DEM} contains the elevations of every cell or the value \textsc{NoData} for cells not part of the DEM. \textbf{At exit,} \textbf{(1)}~\textit{FlowDirs} contains the flow direction of every cell or the value \textsc{NoData} for cells not part of the DEM. \textbf{(2)}~All cells which are part of the DEM have a defined flow direction.}
\label{alg:pf_flowdirs}
\begin{algorithmic}[1]
	\Require \textit{DEM}, \textit{FlowDirs}
	\State Let \textit{Open} be a priority queue with total order
	\State Let \textit{Closed} have the same dimensions as \textit{DEM}
	\State Let \textit{Closed} be initialized to {\sc false}
	\ForAll{$c$ on the edges of \textit{DEM}}
		\State Push $c$ onto \textit{Open} with priority \textit{DEM}($c$)
		\State $\textit{Closed}(c)\gets\textsc{true}$
		\If{$DEM(c)=\textsc{NoData}$}
			\State $\textit{FlowDirs}(c)\gets\textsc{NoData}$
		\Else
			\State $\textit{FlowDirs}(c)$ points off of the DEM
		\EndIf
	\EndFor
	\While{\textit{Open} is not empty}
		\State $c\gets\textsc{pop}(\textit{Open})$
		\ForAll{neighbors $n$ of $c$} \Comment Non-diagonal neighbors first\label{alg:pf_flowdirs_ndn}
			\LineIf{$\textit{Closed}(n)$}{\Continue}
			\If{$\textit{DEM}(n)=\textsc{NoData}$}
				\State $\textit{FlowDirs}(n)\gets\textsc{NoData}$
			\Else
				\State $\textit{FlowDirs}(n)$ points towards $c$
			\EndIf
			\State $\textit{Closed}(n)\gets\textsc{true}$
			\State Push $n$ onto \textit{Open} with priority \textit{DEM}($n$)
		\EndFor
	\EndWhile
\end{algorithmic}
\end{algorithm}

\subsection{Flow Directions}
An alternative to using infinitesimal increments to direct flow is to calculate flow directions directly from the DEM, as in Alg.~\ref{alg:pf_flowdirs}. As in the other Priority-Flood Algorithms, the terrain is flooded inwards from the edges; however, rather than altering cells' elevations, this algorithm always causes the neighbors of the lowest cell to point towards it before pushing them into the priority queue using their unaltered elevations as their priorities. As a result, the algorithm climbs into depressions through their lowest outlet and takes the path of steepest descent to the depressions' minima. Depressions then drain towards their minima and this path, as shown in Fig.~\ref{fig:pf_flowdirs}. All depressions will drain to the edge of the DEM, but elevation information within depressions is still utilized for determining flow directions. The algorithm works well with nested depressions. Alg.~\ref{alg:pf_flowdirs} is, in effect, a depression-carving algorithm: rather than filling depressions, it drills through their walls.

As noted by \citet{Metz2010}, the flow directions determined by this algorithm yield regions of flow accumulation which are located closer to actual rivers than methods which determine flow directions after performing depression-filling.

\subsection{Watershed Labeling}
Watershed labeling applies a common label---such as an integer number---to all cells which drain to a given outlet. The algorithm (Alg.~\ref{alg:barnes1+wl}) for this works in much the same way as the improved Priority-Flood (Alg.~\ref{alg:barnes1}). The DEM is flooded inwards from its edges with the lowest cell always being processed first. Rather than filling cells in depressions, this algorithm merely prioritizes them to the level of their outlet.

Watershed outlets are identified as being unlabeled cells adjacent to a \textsc{NoData} cell. These cells are given a unique label which then floods inwards to cover all the cells in the watershed.

Following execution, watershed boundaries may be identified by locating adjacent cells with differing labels. To mark the border cells, consistently choose either the cell with the lower label, the cell with the higher label, or both.

\begin{algorithm}
\caption{{\sc Improved Priority-Flood+Watershed Labels}: This variation of the {\sc Improved Priority-Flood} follows the work of \citet{BeucherMeyer} and \citet{Beucher2011}. It applies a common label to all cells draining to an outlet. Line~\ref{alg:alsoalter} should be interpreted as pushing a copy of the cell's coordinates into \textit{Pit} with the copy's $z$-value set to $c.z$. If simultaneous watershed labeling and depression-filling is desired, change the original $z$ value of $n$ to $c.z$ before making the copy. \textbf{Upon entry,} \textbf{(1)}~\textit{DEM} contains the elevations of every cell or the value \textsc{NoData} for cells not part of the DEM. \textbf{At exit,} \textbf{(1)}~\textit{Labels} contains a label for every cell or the value \textsc{NoData} for cells not part of the DEM. \textbf{(2)}~All cells which drain to a common point at the edge of the DEM bear the same label.}
\label{alg:barnes1+wl}
\begin{algorithmic}[1]
	\Require \textit{DEM}, \textit{Labels}
	\State Let \textit{Open} be a priority queue
	\State Let \textit{Pit} be a plain queue holding cells' $(x,y,z)$
	\State Let \textit{Labels} have the same dimensions as \textit{DEM}
	\State Let \textit{Labels} be initialized to \textsc{candidate}
	\State $\textit{label}\gets1$
	\ForAll{$c$ on the edges of \textit{DEM}}
		\State Push $c$ onto \textit{Open} with priority \textit{DEM}($c$)
		\State $\textit{Labels}(c)\gets\textsc{queued}$
	\EndFor
	\While{either \textit{Open} or \textit{Pit} is not empty}
		\If{\textit{Pit} is not empty}
			\State $c\gets\textsc{pop}(\textit{Pit})$
		\Else
			\State $c\gets\textsc{pop}(\textit{Open})$
		\EndIf
		\If{$\textit{Labels}(c)=\textsc{queued}$ \aand $\textit{DEM}(c)\ne\textsc{NoData}$}
			\State $\textit{Labels}(c)\gets\textit{label}$
			\State Increment \textit{label}
		\EndIf
		\ForAll{neighbors $n$ of $c$}
			\LineIf{$\textit{Labels}(n)\ne\textsc{candidate}$}{\Continue}
			\State $\textit{Labels}(n)\gets\textit{Labels}(c)$
			\If{$\textit{DEM}(n)\le c.z$}
				\State Push $n$ onto \textit{Pit} with $z=c.z$ \label{alg:alsoalter}
			\Else
				\State Push $n$ onto \textit{Open} with priority \textit{DEM}($n$)
			\EndIf
		\EndFor
	\EndWhile
\end{algorithmic}
\end{algorithm}

\section{Coda}
Special cases and variants of the Priority-Flood Algorithm have been described by many authors. 
This paper generalizes this body of literature (Alg.~\ref{alg:swyfill}) to work optimally on either integer or floating-point data (\textsection\ref{sec:analysis}), as well as on irregular meshes or 4-, 6-, 8-, or $n$-connected grids.

An improvement to the Priority-Flood priority queue (Alg.~\ref{alg:barnes1}) has been described and tested. It runs in $O(m\log_2 m)$ time, where $m\le n$, on floating-point data and in $O(n)$ time on integer data. By comparison, the Planchon--Darboux Algorithm has a time complexity of at least $O(n^{1.2})$ and the generalized Priority-Flood Algorithm has a time complexity of $O(n \log_2 n)$ for floating-point DEMs and $O(n)$ for integer DEMs. Under testing, the new algorithm outperformed the generalized Priority-Flood Algorithm in all cases. Improvements were often greater than 16\% and were as high as 37\%. In addition, a parallel implementation of the Planchon--Darboux Algorithm required upwards of six processors to match the improved Priority-Flood's speed with a single processor.

For its simplicity and speed, Priority-Flood is a good choice. It can fill depressions in such a way that they are guaranteed to drain; it is explicable in 20 lines of pseudocode; and, as demonstrated by the \texttt{C++} reference code (see Supplemental Materials), it can be implemented in fewer than a hundred lines of code.

The Priority-Flood algorithm is also versatile. It can be used to label watersheds (Alg.~\ref{alg:barnes1+wl}) as well as to determine flow directions either by terrain increments (Alg.~\ref{alg:barnes1eps}) or by carving depressions (Alg.~\ref{alg:pf_flowdirs}).

The Supplemental Materials for this paper are available from the journal, as well as at \href{https://github.com/r-barnes/Barnes2013-Depressions}{github.com/r-barnes}. Many of the algorithms presented here are implemented in the {\sc RichDEM} analysis package, available at \url{http://rbarnes.org/richdem} or via email from the authors.

\section{Acknowledgments}
Funding for this work was provided by the Minnesota Environment and Natural Resources Trust Fund under the recommendation and oversight of the Legislative-Citizen Commission on Minnesota Resources (LCCMR). David Mulla was the P.I.\ on this grant. Supercomputing time and data storage were provided by the Minnesota Supercomputing Institute. Adam Clark provided feedback on the paper; Joel Nelson provided LIDAR DEMs and ArcGIS support. In-kind support was provided by Earth Dance Farm, Tess Gallagher, Justin Konen, and Steffi O'Brien.

{\footnotesize
	\bibliographystyle{elsarticle-harv.bst}
	\bibliography{refs}		
}

\end{document}